\documentclass[aps,eqsecnum,nofootinbib,showpacs,twocolumn]{revtex4}
\usepackage{amsmath,amsfonts,amssymb,bm}
\usepackage[dvips]{graphicx}
\usepackage{color}
\usepackage{wrapfig}
\usepackage{graphicx}
\usepackage{color}  
\def\be{\begin{eqnarray}}
\def\ee{\end{eqnarray}}
\def\bec{\begin{center}}
\def\eec{\end{center}}

\def\p{\partial}
\bmdefine{\bmk}{\bm{k}}
\bmdefine{\bmx}{\bm{x}}
\newcommand{\IP}[2]{\langle~#1~\vert~#2~\rangle}

\begin{document}
\title{The final fate of instability of Reissner-Nordstr\"{o}m-anti-de Sitter 
black holes by charged complex scalar fields}
\author{Kengo Maeda}
\email{maeda302@sic.shibaura-it.ac.jp}
\affiliation{Faculty of Engineering,
Shibaura Institute of Technology, Saitama, 330-8570, Japan}

\author{Jun-ichirou Koga}
\email{koga@gravity.phys.waseda.ac.jp}
\affiliation{Advanced Research Institute for Science and Engineering, Waseda University, Shinjuku, 
Tokyo 169-8555}

\author{Shunsuke Fujii}
\email{fujiis@sic.shibaura-it.ac.jp}
\affiliation{Faculty of Engineering,
Shibaura Institute of Technology, Saitama, 330-8570, Japan}
\date{\today}
\begin{abstract}
We investigate instability of $4$-dimensional Reissner-Nordstr\"{o}m-anti-de 
Sitter~(RN-AdS$_4$) black holes with various topologies by charged scalar field perturbations. 
We numerically find that the RN-AdS$_4$ black holes become unstable against the linear 
perturbations below a critical temperature. It is analytically shown that charge extraction 
from the black holes occurs during the unstable evolution. 
To explore the end state of the instability, we perturbatively construct static black hole 
solutions with the scalar hair near the critical temperature. It is numerically found that the entropy 
of the hairly black hole is always larger than the one of the unstable RN-AdS$_4$ black hole in the 
microcanonical ensemble. 
Our results support the speculation that the black hole with charged scalar hair always 
appears as the final fate of the instability of the RN-AdS$_4$ black hole. 
\end{abstract}

\maketitle

\section{Introduction}
Motivated by a holographic model of superconductors, there has been renewal 
of interest in the Einstein-Maxwell-scalar system in asymptotically anti-de Sitter spacetime. 
The model is constructed by a gravitational theory of a charged complex scalar field coupled to 
the Maxwell field~(for review, see Refs.~\cite{Hartnoll:2009sz,Herzog:2009xv, Horowitz:2010rv}) 
via AdS/CFT~(anti-de Sitter/conformal field theory) duality~\cite{Maldacena:1997re}. 
According to the AdS/CFT dictionary,  a condensate in the holographic superconductor is   
described by a hairly black hole dressed with the charged scalar field.   
It was numerically found that the hairly black hole exists below a critical temperature $T_c$ in 
the plane-symmetric spacetime and the phase transition is second-order~\cite{HHH2008verII}.  

For a gravity system minimally coupled with a neutral scalar field, it is well known that 
the scalar field perturbation with mass satisfying  Breitenlohner-Freedman~(BF) bound 
does not cause any instabilities of AdS spacetime~\cite{BF}. 
So, even though a black hole solution with the scalar hair exists in asymptotically 
AdS spacetime~\cite{tmn}, Schwarzschild-AdS~(Sch-AdS) spacetime should be stable against the 
scalar field perturbations if BF bound is satisfied. In $4$ and $5$-dimensional $N=8$ gauged 
supergravities, this speculation was supported by the numerical calculation that the entropy of 
the Sch-AdS black hole is larger than the one of the hairly black hole in the microcanonical 
ensemble~\cite{hertogmaeda}. 
In this paper we investigate how this picture is modified in the gravity system minimally coupled with 
a charged scalar field with mass satisfying BF bound.
As discussed in Ref.~\cite{gubser77}, RN-AdS black hole with low temperature may be unstable against 
the perturbation of the charged scalar field since the effective mass of the scalar field is 
negative near the horizon. So, we naively expect that a black hole with the scalar hair appears as 
the final fate of the instability even though BF bound is satisfied. 

In AdS spacetime, the Dirichlet boundary condition is usually imposed at spatial infinity when we consider the 
dynamical evolution. Under the boundary condition, energy flux and charge current do not enter through the 
infinity, and therefore, the total mass and charge of the spacetime are conserved during the dynamical evolution. 
So, if RN-AdS$_4$ black hole evolves to the hairly black hole as the end state of the instability, 
the entropy of the hairly black hole should be larger than the one of RN-AdS$_4$ black hole in the microcanonical 
ensemble due to the second law of black hole. 

To explore the above speculation we first investigate instability of RN-AdS$_{4}$ black holes with 
various topologies of horizon by charged complex scalar fields with various mass satisfying 
BF bound. 
Second, we numerically show that the entropy of the hairly black hole is always larger 
than the one of RN-AdS$_{4}$ black hole whenever the RN-AdS$_{4}$ black hole is unstable. 
An analytic calculation also shows that charge is extracted from the unstable black hole via 
the charged scalar field.  

The plan of our paper is as follows: In Sec.~II, we investigate the lowest quasi-normal~(QN) 
frequencies of RN-AdS$_{4}$ black holes both numerically and analytically near the second order phase 
transition. In Sec.~III, we construct the hairly black hole solutions by solving the field equations 
perturbatively near the phase transition. The entropy of the hairly black hole is numerically calculated 
by using the first law of black hole. Conclusion and discussions are devoted to Sec.~IV. 

\section{Quasi-normal frequencies near the second order phase transition}

We consider the Lagrangian density of Einstein-Maxwell-scalar system in $4$-dimensional 
asymptotically anti-de Sitter~(AdS$_4$) spacetime:     
\begin{align}
\label{lagrangian-density}
& {\cal L}=R+\frac{6}{L^2}-\frac{F^{ab}F_{ab}}{4}-m^2|\psi|^2-|D\psi|^2, \nonumber \\
& D_a=\nabla_a-iqA_a, \qquad F_{ab}:=\p_aA_b-\p_bA_a 
\end{align}
where $L$, $m$, and $q$ are the AdS radius, the mass, and the charge of the complex scalar field, 
respectively. The equations of motion are given by 
\begin{subequations}
\begin{align}
& D_aD^a\psi=m^2\psi, 
\label{eq-scalar} \\
& \nabla_a{F_b}^a=j_b, \qquad j_a=iq\,[(D_a\psi)^\dagger\psi-(D_a\psi)\psi^\dagger], 
\label{eq-gauge} \\
& G_{ab}=\frac{3g_{ab}}{L^2}+\frac{1}{2}F_{ac}{F_b}^c-\frac{g_{ab}}{8}F^2-\frac{m^2}{2}|\psi|^2g_{ab}
\nonumber \\
& -\frac{g_{ab}}{2}|D\psi|^2+\frac{1}{2}
[D_a\psi(D_b\psi)^\dagger+(D_a\psi)^\dagger D_b\psi]. 
\label{eq-einstein} 
\end{align}
\label{eq-motion}
\end{subequations}
The simple solution of the equations are given by Reissner-Nordstr\"{o}m~(RN)-AdS$_4$ 
black hole solutions with various curvature of horizon. 
The metric is given by 
\begin{subequations}
\begin{align}
& ds^2=-f(r)dt^2+\frac{dr^2}{f(r)}+r^2d\Omega_2^2, 
\label{eq:line_element} \\
& f(r)=k-\frac{2M}{r}+\frac{\rho^2}{4r^2}+\frac{r^2}{L^2} \nonumber \\
&=\frac{1}{L^2r^2}(r-r_+)(r-r_-)(r^2+ar+b), 
\label{eq:f0} \\
&  a:=r_++r_-, \qquad b:=kL^2+(r_++r_-)^2-r_+r_-, \nonumber \\
& T=\frac{(r_+-r_-)(r_+^2+ar_++b)}{4\pi L^2r_+^2}, 
\label{temperature}
\end{align}
\label{RN-AdS-sol}
\end{subequations}
where $d\Omega_2^2$, $r_+~(r_-)$, and $T$ are the metric of 2-dimensional Einstein space 
with constant scalar curvature 
$2k~(k=0,\,\pm 1)$, the outer~(inner) horizon radius, and the Hawking temperature, respectively. 
The mass~(charge) parameter $M$~($\rho)$ corresponds to the total mass~(charge) for $k=1$ and 
the mass~(charge) density for $k=0,\,-1$, respectively. 
These two parameters are rewritten in terms of $r_\pm$, $L$, and $k$ as
\begin{subequations}
\begin{align}
\label{def-rho}
 \rho=2\sqrt{\frac{r_+r_-(kL^2+(r_++r_-)^2-r_+r_-)}{L^2}}, 
\end{align}
\begin{align}
\label{def-mass}
M=\frac{r_++r_-}{2L^2}(kL^2+r_+^2+r_-^2), 
\end{align}
\label{def-rho-mass}
\end{subequations}
and the gauge potential $A_\mu$ is given by
\begin{align}
\label{eq:gauge-potential(0)} 
 A_\mu dx^\mu=\Phi(r)dt=\rho\left(\frac{1}{r_+}-\frac{1}{r}\right)dt. 
\end{align}

We obtain the lowest quasi-normal~(QN) frequencies of RN-AdS$_4$ black holes by considering 
perturbation of the solutions. 
Since $\psi=0$ on the RN-AdS$_4$ solutions, the scalar perturbation decouples from the 
perturbations of the other fields. So, we solve Eq.~(\ref{eq-scalar}) for the fixed metric~(\ref{eq:line_element}) 
and the gauge potential~(\ref{eq:gauge-potential(0)}). Near the horizon, $\psi$ behaves as 
\begin{align}
\psi_\mp\sim e^{-i\omega t}(1-u)^{\mp\gamma \omega i}, \,\, 
\gamma=\frac{L^2r_+^2}{(r_+-r_-)(r_+^2+ar_++b)}, 
\end{align}
where $u$ is defined by $u:=r_+/r$. From the retarded condition, we must impose 
the ``ingoing wave" boundary condition: $\psi\sim \psi_-$ on the horizon.  
Defining a new variable $\Psi$ as 
\begin{align}
\psi(t,u)=e^{-i\omega t}(1-u)^{-\gamma \omega i}\Psi(u), 
\end{align}
Eq.~(\ref{eq-scalar}) is written as
\begin{align}
\label{eq-Psi}
& \Psi''(u)+\left(\frac{2\gamma\omega i}{1-u}+\frac{f'(u)}{f(u)}\right)\Psi'(u) \nonumber \\
& +\Biggl[\frac{\gamma\omega i}{1-u}\left(\frac{f'(u)}{f(u)}+\frac{\gamma\omega i+1}{1-u}\right)
\nonumber \\
&+\frac{r_+^2}{u^4f^2(u)}(\omega+qA_t)^2-\frac{r_+^2m^2}{u^4f(u)}\Biggr]\Psi(u)=0, \\
& f(u)=\frac{r_+^2}{L^2u^2}(1-u)\left(1-\frac{r_-}{r_+}u\right)
\left(1+\frac{a}{r_+}u+\frac{b}{r_+^2}u^2 \right),  \nonumber
\end{align}
where a prime denotes the derivative with respect to $u$. In terms of this variable $\Psi$, 
the ``ingoing wave" boundary condition is described by the regularity condition:
\begin{align}
\label{bn-regular}
\Psi(u)\, \mbox{is regular at}\, u=1. 
\end{align}

Near the spatial infinity, $\Psi$ behaves as 
\begin{align}
\Psi(u)=c_1 u^{\Delta_+}+c_2 u^{\Delta_-}, \,\, \Delta_\pm=\frac{3}{2}\pm\sqrt{9/4+L^2m^2}, 
\end{align} 
where Breitenlohner-Freedman~(BF) bound~\cite{BF} is represented by $m^2\ge -9/4L^2$. Since we 
are concerned with the time evolution under which both the mass~(or mass density) $M$ and the 
charge~(or charge density) $\rho$ are conserved, 
we shall impose Dirichlet boundary condition at infinity~\footnote{When $-9/4L^2\le m^2<-5/4L^2$, there 
are two normalizable modes. So, imposing $c_1=0$ instead of $c_2=0$ is also possible as the 
boundary condition. It is well known, however, that 
the boundary condition in which neither $c_1$ nor $c_2$ vanishes is unphysical since 
the spacetime is unstable~\cite{hertoghorowitz}.}:
\begin{align}
\label{bn-Dirichlet}
\Psi(u)\sim u^{\Delta_+}. 
\end{align}

According to the critical slowing down phenomenon, the lowest QN frequencies $\omega$ should 
approach zero toward the second order phase transition $T=T_c$~\cite{mno2008, mno2009}. This implies that one can 
expand the solution of Eq.~(\ref{eq-Psi}) as
\begin{align}
\Psi(u)=\Psi_0(u)+\omega \Psi_1(u)+O(\omega^2) 
\end{align}
near the phase transition, where $\Psi_0(u)$ is the marginally stable solution. 
Denoting the deviations from the phase transition as 
\begin{align}
& r_+=r_c+\delta r_+, \nonumber \\
& \gamma=\gamma_c+\delta \gamma, \nonumber \\
& f(u)=f_c(u)+\delta f(u), \nonumber \\
& \Phi(u)=\Phi_c(u)+\delta \Phi(u),  
\end{align}
we write down the equations for $\Psi_0(u)$ and $\Psi_1(u)$ as
\begin{align}
\label{eq-Psi0}
{\cal L}\Psi_0(u)=0, 
\end{align}
\begin{align}
\label{eq-Psi1}
{\cal L}(\omega \Psi_1)=j_1(u), 
\end{align}
where the differential operator ${\cal L}$ and the source term $j_1$ are defined by 
\begin{align}
{\cal L}:=\frac{d^2}{du^2}+\frac{f_c'(u)}{f_c(u)}\frac{d}{du}+
\frac{r_{c}^2}{u^4f_c(u)}\left(\frac{q^2\Phi_c^2(u)}{f_c(u)}-m^2\right), 
\end{align}
\begin{subequations}
\begin{align}
& \mbox{Re}[j_1]:=\frac{2\gamma_c\omega_I}{1-u}\Psi_0'(u)
-\delta\left(\frac{f'(u)}{f(u)}\right)\Psi_0'(u) \nonumber \\
& +\Biggl[\frac{\gamma_c\omega_I}{1-u}\left(\frac{f_c'(u)}{f_c(u)}+\frac{1}{1-u}\right)
-\frac{2q^2r_c\Phi_c^2(u)}{u^4f_c(u)}\delta \left(\frac{r_+}{f(u)}\right)  
\nonumber \\
&-\frac{2q\omega_Rr_c^2}{u^4f_c^2(u)}\Phi_c(u)-
\frac{2q^2r_c^2\Phi_c(u)\delta \Phi(u)}{u^4f_c^2(u)} \nonumber \\
& +\frac{2m^2r_c\delta r_+}{u^4f_c(u)} 
-\frac{m^2r_c^2\delta f(u)}{u^4f_c^2(u)}\Biggr]\Psi_0(u), 
\end{align}
\begin{align}
& \mbox{Im}[j_1]:=-\frac{\gamma_c\omega_R}{1-u}\left[2\Psi_0'(u)+\left(\frac{f_c'(u)}{f_c(u)}
+\frac{1}{1-u}\right)\Psi_0(u)\right] \nonumber \\
& -\frac{2q\omega_Ir_c^2}{u^4f_c^2(u)}\Phi_c(u)\Psi_0(u) 
\end{align}
\label{eq-j1}
\end{subequations}
for $\omega=\omega_R+i\omega_I$. To derive Eq.~(\ref{eq-j1}), we assumed that $\Psi_0$ 
is a real function of $u$ without loss of generality. 

To derive $\omega$ from Eq.~(\ref{eq-Psi1}), let us define the inner product as 
\begin{align}
\IP{\psi_a}{\psi_b}:=\int^1_0\psi_a^\dagger(u)\psi_b(u)du 
\end{align} 
for any functions $\psi_a$ satisfying both boundary conditions~(\ref{bn-regular}) and 
(\ref{bn-Dirichlet}). Under these boundary conditions, the differential operator 
$\hat{{\cal L}}:=f_c(u){\cal L}$ is shown to be an Hermite operator and by Eq.~(\ref{eq-Psi0}) we 
obtain  
\begin{align}
\label{co-hermite}
\IP{\hat{j}_1}{\Psi_0}=\IP{\hat{\cal L}\Psi_1}{\Psi_0}=\IP{\Psi_1}{\hat{\cal L}\Psi_0}=0, 
\end{align}  
where $\hat{j}_1(u):=f_c(u)j_1(u)/\omega$. 
The real and imaginary parts of Eq.~(\ref{co-hermite}) yield the lowest QN frequencies $\omega$ as
\begin{subequations}
\begin{align}
\label{real-omega}
\omega_R=-\frac{2qr_c\omega_I}{\Psi_0^2(1)}\int^1_0\frac{\Phi_c(u)\Psi_0^2(u)}{u^4f_c(u)}du, 
\end{align}
\begin{align}
\label{imaginary-omega}
& \omega_Ir_{c}\left[\Psi_0^2(1)+\frac{4q^2r_{c}^2}{\Psi_0^2(1)}
\left(\int^1_0\frac{\Phi_c(u)\Psi_0^2(u)}{u^4f_c(u)}du\right)^2\right] \nonumber \\
& =\frac{1}{2}\int^1_0\delta\left(\frac{f'(u)}{f(u)}\right)f_c(u)\{\Psi^2_0(u)\}'du \nonumber \\
&+2q^2r_c\int^1_0\frac{\Psi_0^2(u)}{u^4}
\delta\left(\frac{r_+}{f(u)}\right)\Phi_c^2(u)du \nonumber \\
& +\int^1_0\frac{\Psi_0^2(u)}{u^4f_c(u)}\left[q^2r_c^2\delta(\Phi^2(u))-m^2
f_c^2(u)\delta\left(\frac{r_+^2}{f(u)}\right)\right]du. 
\end{align}
\label{omega}
\end{subequations}
It is noteworthy that $\omega$ is independent of the absolute amplitude of $\Psi_0(1)$ and 
it is determined by solving only the leading order equation~(\ref{eq-Psi0}). 

Hereafter, we shall fix $L$ to $L=1$. Then, from Eqs.~(\ref{temperature}) and (\ref{def-rho-mass}), 
$r_\pm$ and $T$ are functions of $M$ for given $k$ and $\rho$. So, for the sake of later 
convenience, we shall define the deviation parameter from the second order phase transition as
\begin{align}
\label{def-epsilon}
\epsilon:=\frac{M_c-M}{M_c}, 
\end{align}
where $M_c$ is the mass at the critical temperature $T=T_c$. 
Since $M$ is locally a smooth function of $k$, $\rho$, and $T$, 
$\epsilon$ approaches zero in the limit $T\to T_c$. 
From Eqs.~(\ref{def-rho-mass}) $\delta r_{\pm}$ is related to $\epsilon$ as 
\begin{subequations}
\begin{align}
\label{deviation-}
\delta r_-=-\frac{r^*_{c}(kL^2+3r_{c}^2+2r_{c}r^*_{c}+{r^*_{c}}^2)}
{r_{c}(kL^2+3{r^*_{c}}^2+2r_{c}r^*_{c}+r_{c}^2)}\delta r_+
\end{align}
and  
\begin{align}
\delta r_+=-\frac{(r_{c}+r^*_{c})(kL^2+r_{c}^2+{r^*_{c}}^2)\epsilon}
{(kL^2+3r_{c}^2+2r_{c}r^*_{c}+{r^*_{c}}^2)(1-r^*_{c}/r_{c})},  
\end{align}   
\label{relation}
\end{subequations}
where $r^*_{c}$ is the inner horizon radius $r_-$ at $T=T_c$. 
To derive Eq.~(\ref{relation}), we impose the condition that $\rho$ is constant 
along the deviation.   

For a generic value of $M$ both of the two boundary conditions~(\ref{bn-regular}) and (\ref{bn-Dirichlet}) 
are not necessarily satisfied, as Eq.~(\ref{eq-Psi0}) is the second order linear differential equation. 
By using Mathematica, we numerically scan through possible values of $M$ until the two boundary 
conditions are satisfied and find $M_c$ for given $m$, $q$, $k$, and $\rho$. 
Then, representing $\delta f$, $\delta\Phi$, and 
$\delta r_+$ by $\epsilon$ via Eqs.~(\ref{relation}) and integrating Eqs.~(\ref{omega}) 
numerically, we obtain $\omega_I$.  
\begin{figure}
\includegraphics[width=8.0truecm,clip]{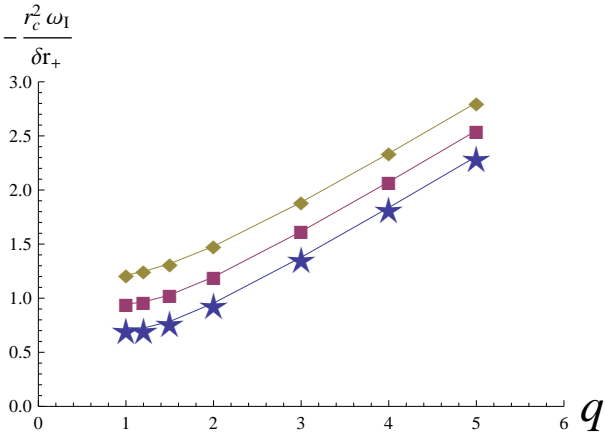}
\caption{ \label{fig:QNMs}
(color online). The $q$-dependence of $-r_c^2\omega_I/\delta r_+$ is shown for various $k$ 
in the case of $m^2=-2/L^2$ and $\rho=2$. The star, box, and diamond 
represent $k=1$, $k=0$, and $k=-1$, respectively. }
\end{figure}
\begin{figure}
\includegraphics[width=8.0truecm,clip]{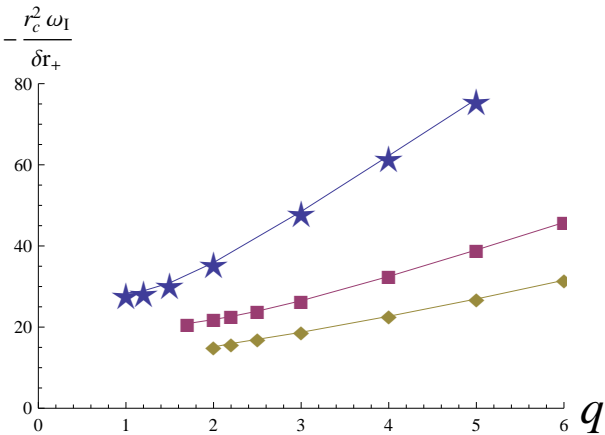}
\caption{ \label{fig:QNMs1}
(color online). The $q$-dependence of $-r_c^2\omega_I/\delta r_+$ is shown for several $m^2$ 
in the case of $k=0$ and $\rho=60$. The star, box, and diamond 
represent $m^2=-2/L^2$, $m^2=0$, and $m^2=7/4L^2$, respectively. At the left edge of each plot, 
the RN-AdS$_4$ black holes become almost extremal.}
\end{figure}

Since all the deviations are proportional to $\epsilon$, $\omega$ is also proportional to $\epsilon$, 
in agreement with the critical slowing down~\cite{mno2008,mno2009}. 
We find that $\omega_I$ becomes positive~(negative) for $\epsilon>0~(<0)$, independent of $k$, $q$, and $m^2$.   
In Fig.~1 we plot $q$-dependence of the dimensionless quantity $-r_c^2\omega_I/\delta r_+$ 
for various $k$ in the case of $m^2=-2/L^2$ 
and $\rho=2$. In Fig.~2, we plot $q$-dependence of $-r_c^2\omega_I/\delta r_+$ 
for several $m^2$ in the case of $\rho=60$. 
These figures show that RN-AdS$_4$ black holes become unstable against the 
linear perturbation of the complex scalar field when $T<T_c~(\epsilon>0)$,~\footnote{We assumed 
that $\epsilon>0$ for $T<T_c$. This corresponds to the assumption that the specific heat of the 
RN-AdS$_4$ black hole is positive at $T=T_c$.} as $\epsilon/\delta r_+<0$, independent of $k$ and $m^2$.  

The charge current through the horizon is calculated by contracting $j_a$ in Eq.~(\ref{eq-gauge}) 
with the killing vector $k^a=(\p_t)^a$, which is null on the horizon. By Eq.~(\ref{real-omega}), we obtain 
\begin{align}
\label{current}
j_ak^a|_{r=r_c}=4q^2r_c\omega_Ie^{2\omega_I t}\int^1_0\frac{\Phi_c(u)\Psi_0^2(u)}{u^4f_c(u)}du, 
\end{align}
which shows that $j_ak^a|_{r=r_c}$ is positive when $\omega_I>0$. This implies that charge extraction occurs 
from the RN-AdS$_4$ black hole when it is unstable.

\section{The entropy for hairly black holes}
In this section, we investigate the final fate of the instabilities found in Sec.~II by  
calculating the entropy of the hairly black holes. We start by constructing 
the static hairly black hole solutions near the phase transition. Near the phase transition, 
the equations of motion~(\ref{eq-motion}) can be perturbatively solved since the scalar 
field $\psi$ is very small. To solve the perturbed equations easily, we adopt a coordinate 
system $\{T,\,y \}$ where the location of the horizon is set to be $y=1$. 
So, we shall take the following ansatz for the metric and the gauge field:
\begin{subequations}
\begin{align}
\label{metric-ansatz}
ds^2=-g(y)dT^2+\frac{dy^2}{g(y)}+R^2(y)d\Omega_2^2,  
\end{align}
\begin{align} 
\label{gauge-ansatz}
A_\mu dx^\mu=\phi(y)dT.  
\end{align}
\label{ansatz}
\end{subequations}
Assuming that $\psi$ is real without loss of generality, the scalar and Maxwell 
equations~(\ref{eq-scalar}) and (\ref{eq-gauge}) are reduced to the 
two coupled differential equations:
\begin{subequations}
\begin{align} 
\label{eq-scalar-static}
g\psi''+\left(\frac{2gR'}{R}+g'\right)\psi'+\frac{q^2\phi^2\psi}{g}=m^2\psi, 
\end{align}
\begin{align}
\label{da-gauge-static}
\phi''+\frac{2R'}{R}\phi'=\frac{2q^2\phi\psi^2}{g}, 
\end{align}
\label{eq-two-coupled}
\end{subequations}
where a prime denotes the derivative with respect to $y$. The metric functions 
$g$ and $R$ are obtained by solving 
the Einstein equations~(\ref{eq-einstein}): 
\begin{subequations}
\begin{align}
\label{static-dev}
\frac{R''}{R}=-\frac{q^2\phi^2\psi^2}{2g^2}-\frac{\psi'^2}{2}, 
\end{align}
\begin{align}
\label{static-constraint}
& \frac{R'}{R}g'=\frac{k}{R^2}-\frac{gR'^2}{R^2}+\frac{3}{L^2}-\frac{1}{4}\phi'^2
-\frac{1}{2}m^2\psi^2 \nonumber \\
& +\frac{1}{2}g\psi'^2+\frac{q^2}{2g}\phi^2\psi^2. 
\end{align}
\label{static-einstein}
\end{subequations}

As shown in Refs.~\cite{maedaokamura2008,mno2010}, we can expand the functions $\psi$, $g$, $R$, $\phi$, 
and temperature $T$ near the second order phase transition as
\begin{align}
\label{expansion-T}
& \psi(y)=\epsilon^{1/2}\psi_1(y)+\epsilon^{3/2}\psi_2(y)+\cdots, \nonumber \\
& g(y)=g_c(y)+\epsilon\,g_1(y)+\cdots, \nonumber \\
& R(y)=r_{c}y+\epsilon\,R_1(y)+\cdots, \nonumber \\
& \phi(y)=\phi_c(y)+\epsilon\,\phi_1(y)+\cdots, \nonumber \\
& T(\epsilon)=T_c+\epsilon T_1+\cdots. 
\end{align} 
The zeroth order solutions $g_c(y)$ and $\phi_c(y)$ are simply given by the coordinate transformation: 
\begin{align}
r=r_{c}\,y, \qquad t=T/r_{c} 
\end{align}
of the RN-AdS$_4$ black hole solution~(\ref{RN-AdS-sol}) as
\begin{subequations}
\begin{align}
\label{sol-gc}
g_c(y)=\frac{1}{L^2y^2}(y-1)(y-r^*_{c}/r_{c})\left(y^2+\frac{a}{r_{c}}y+\frac{b}{r_{c}^2}\right), 
\end{align}
\begin{align}
\label{sol-gauge-T}
\phi_c(y)=\frac{\rho}{r_{c}^2}\left(1-\frac{1}{y}\right).  
\end{align}
\end{subequations}
Then, $\psi_1(y)$, $\phi_1(y)$, $R_1(y)$, and $g_1(y)$ obey the following linear differential 
equations:
\begin{align}
\label{eq-psi1}
\frac{\left(y^2g_c\psi_1'\right)'}{y^2}={\cal M}^2\psi_1, 
\end{align}
\begin{align}
\label{eq-phi1}
\phi_1''+\frac{2}{r_{c}}\left(\frac{R_1'}{y}-\frac{R_1}{y^2}  \right)\phi_c'
+\frac{2}{y}\phi_1'=\frac{2q^2\phi_c\psi_1^2}{g_c}, 
\end{align}
\begin{align}
\label{eq-R1}
R_1''=-\frac{r_{c}y}{2}\left(\frac{q^2\phi_c^2\psi_1^2}{g_c^2}+\psi_1'^2 \right), 
\end{align}
\begin{align}
\label{eq-g1}
& \frac{1}{y}g_1'+\frac{g_1}{y^2}+\frac{1}{r_{c}}\left(g_c'+\frac{2g_c}{y}\right)
\left(\frac{R_1'}{y}-\frac{R_1}{y^2}\right)g_c'
+\frac{2kR_1}{r_{c}^3y^3} \nonumber \\
& =-\frac{1}{2}\phi_c'\phi_1'
-\frac{m^2}{2}\psi_1^2+\frac{1}{2}g_c\psi_1'^2+\frac{q^2}{2g_c}\phi_c^2\psi_1^2,  
\end{align}
where ${\cal M}$ is the effective mass defined by 
\begin{align}
\label{effective-mass}
{\cal M}^2(y):=m^2-q^2\phi_c^2/g_c 
\end{align}
which can be negative~\cite{gubser77}.  

The solutions of Eqs.~(\ref{eq-phi1}) and (\ref{eq-R1}) are formally given by 
\begin{subequations}
\begin{align}
\label{sol-phi1}
 \phi_1'=\frac{2}{y^2}\int^y_\infty
\left[\frac{q^2\phi_c\psi_1^2z^2}{g_c}+\frac{\phi_c'}{r_{c}}(R_1-z R_1')\right]dz+\frac{c_3}{y^2}, 
\end{align}
\begin{align}
\label{sol-R1}
& R_1(y)={\cal R}(y)+c_1y+c_2, \\
& {\cal R}(y):=
-\frac{r_c}{2}\int^y_\infty \int^{z}_\infty
\left(\frac{q^2\phi_c^2(w)\psi_1^2(w)}
{g_c^2(w)}+\psi_1'^2(w) \right)wdwdz,   
\end{align}
\label{sol-1T}
\end{subequations}
where $c_1$, $c_2$, and $c_3$ are integral constants. Note that $\psi_1(y)$ is given by $\Psi_0(1/y)$, as 
Eq.~(\ref{eq-psi1}) is equivalent to Eq.~(\ref{eq-Psi0}) by setting $y=1/u$. 
So, all solutions are determined by $\Psi_0$ except the integral constants.  

Since $\psi_1$ rapidly decays as $\psi=O(y^{-\lambda})~(\lambda>3/2)$ near infinity, the first term of 
Eq.~(\ref{sol-phi1}) decays as $O(y^{-3})$, which does not contribute to the charge~(or charge density) 
$\rho_h$ of the hairly black hole solutions. $\rho_h$ is given by 
\begin{align}
\rho_h=\rho\left(1+\frac{2c_1\epsilon}{r_c}+\frac{r_c^2c_3\epsilon}{\rho}\right) 
\end{align}
up to the order $O(\epsilon)$. So, equating $\rho_h=\rho$, we obtain the constant $c_3$ as
\begin{align}
\label{constant-c3}
c_3=-\frac{2c_1\rho}{r_c^3}. 
\end{align}
 
The formal solution of Eq.~(\ref{eq-g1}) is given by
\begin{align}
\label{sol-g1}
& g_1=\frac{1}{r_cy}\int^y_1(R_1-R_1'y)\left(g_c'+\frac{2g_c}{y}\right)dy
-\frac{2}{y}\int^y_1\frac{kR_1}{r_c^3y}dy \nonumber \\
& -\frac{1}{2y}\int^y_1y^2\phi_c'\phi_1'dy 
-\frac{m^2}{2y}\int^y_1 y^2 \psi_1^2dy \nonumber \\
& +\frac{1}{2y}\int^y_1 y^2 g_c {\psi'}_1^2dy 
+\frac{q^2}{2y}\int^y_1\frac{y^2\phi_c^2\psi_1^2}{g_c}dy. 
\end{align}

To read off the mass (or mass density) $M$ from the metric~(\ref{metric-ansatz}), we need to 
take the coordinate transformation $T\to t$: 
\begin{align}
T=t(r_{c}+c_1\epsilon)
\end{align}
so that the leading term of the metric~(\ref{metric-ansatz}) 
near the spatial infinity $y\to \infty$ coincides with 
the asymptotic form of the metric~(\ref{RN-AdS-sol}) of RN-AdS$_4$ black holes: 
\begin{align}
ds^2= -\frac{R^2}{L^2}dt^2+\frac{L^2dR^2}{R^2}+R^2d\Omega_2^2. 
\end{align}
Substituting Eqs.~(\ref{sol-1T}) and~(\ref{constant-c3}) into 
Eq.~(\ref{sol-g1}), we find that the asymptotic fall-off of $-g_{tt}$ component behaves as
\begin{align}
\label{asmp-tt}
& -g_{tt}=f_c(R)+\frac{2\epsilon\chi(\psi_1)}{R}
-\frac{4\pi(c_1+c_2)r_cT_c}{R}\epsilon \nonumber \\
&+\cdots+O(\epsilon^2) \nonumber \\
&=\frac{R^2}{L^2}+k-\frac{2M}{R}+\cdots+O(\epsilon^2).  
\end{align}
Here, $\chi$ is a complicated function of $\psi_1$ and it vanishes when $\psi_1=0$. 
By Eqs.~(\ref{def-epsilon}) and (\ref{asmp-tt}), 
the combination of the constants $c_1+c_2$ is rewritten by $\chi$ and $M_c$ as
\begin{align}
\label{c1+c2}
c_1+c_2=\frac{\chi(\psi_1)-M_c}{2\pi T_cr_c}. 
\end{align}
Note that RN-AdS$_4$ black hole solutions with mass~(or mass density) $M=M_c(1-\epsilon)$ are automatically 
derived by setting $\psi_1=\chi(\psi_1)=0$. 

We calculate the entropy~(or entropy density) $S$ of the hairly black holes with the same mass~(or mass density) $M$ by 
using the first law~(\ref{eqn:FirstLaw}) for a fixed charge~(or charge density) $\rho$ near the 
critical temperature $T_c$: 
\begin{align}
\frac{\delta S}{\delta M}(\epsilon)=\frac{1}{T(\epsilon)}=
\frac{1}{T_c}-\frac{T_1}{T_c^2}\epsilon+O(\epsilon^2). 
\end{align}
Integrating this differential equation under the relation 
$\delta M=-M_c\,\delta\epsilon$, we obtain 
\begin{align}
\label{entropy}
& S=S_c+S_1\epsilon+S_2\epsilon^2+\cdots \nonumber \\
& =S_c-\frac{M_c\epsilon}{T_c}+\frac{M_cT_1\epsilon^2}{2T_c^2}+O(\epsilon^3). 
\end{align}
This means that the difference of the entropy~(or entropy density) between RN-AdS$_4$ black hole and the hairly black hole 
appears at $O(\epsilon^2)$:
\begin{align}
& \Delta S=\frac{\epsilon^2M_c}{2T_c^2}\Delta T_1=S_{hairly BH}-S_{RN-AdS_4 BH} \nonumber \\
&\propto R^2(1)_{hairly BH}-R^2(1)_{RN-AdS_4 BH}=O(\epsilon^2), 
\end{align}
and then ${R_1(1)}|_{hairly BH}={R_1(1)}|_{RN-AdS_4 BH}$. Hence, by Eqs.~(\ref{sol-R1}) and 
(\ref{c1+c2}), $c_1+c_2$ is rewritten by ${\cal R}$ as
\begin{align}
\label{c1+c2-1}
& c_1+c_2=-{\cal R}(1)-\frac{M_c}{2\pi T_cr_c}. 
\end{align}
As a check of numerical calculation below, we verified that the r.h.s. of Eq.~(\ref{c1+c2}) agrees with the r.h.s. of 
Eq.~(\ref{c1+c2-1}) within the numerical error. 

We obtain the Hawking temperature $T$ from the regularity of the Euclidean solution analytically 
continued by $t\to i\tau$ as
\begin{align}
T(\epsilon)=T_c+\frac{r_{c}}{4\pi}\left(g_1'(1)+\frac{c_1}{r_{c}}g_c'(1)\right)\epsilon
+O(\epsilon^2), 
\end{align}
where the critical temperature $T_c$ is given by $T_c=r_{c}g_c'(1)/4\pi$. 
By using Eqs.~(\ref{sol-1T}), (\ref{sol-g1}), and 
(\ref{c1+c2-1}), $T_1$ in Eq.~(\ref{expansion-T}) is expressed by  
\begin{align}
& T_1=\frac{r_c}{4\pi}\left(g_1'(1)+\frac{c_1}{r_c}g_c'(1)\right) \nonumber \\
&=\frac{1}{4\pi}\Biggl[-\frac{r_{c}g_c'(1)}{2}\int^\infty_1
\left(\frac{q^2\phi_c^2(y)\psi_1^2(y)}{g_c^2(y)}+\psi_1'^2(y)\right)ydy \nonumber \\
&+\frac{\rho}{r_{c}}\int^\infty_1\frac{q^2\phi_c(y)\psi_1^2(y)y^2}{g_c(y)}dy
-\frac{m^2}{2}r_c\psi_1^2(1)\Biggr] \nonumber \\
&-\frac{M_c}{8\pi^2T_cr_c^3}\left(4\pi T_cr_c+\frac{\rho^2}{r_c^2} \right). 
\end{align}
The first term in the second equality corresponds to the difference of $T_1$ 
between the RN-AdS$_4$ black hole and the hairly one. Eliminating $\psi_1'^2$ term by 
Eq.~(\ref{eq-psi1}) and using integration by parts, we finally obtain a useful expression 
for the difference $\Delta T_1:=T_1|_{hairly BH}-T_1|_{RN-AdS}$ as  
\begin{align}
\label{Delta-T}
& \frac{\Delta T_1}{T_c}=\frac{\psi_1^2(1)}{4}+\frac{1}{2}
\int^\infty_1
\left(m^2\psi_1-g_c'\psi_1'\right)\frac{\psi_1}{g_c} ydy \nonumber \\
& +\frac{1}{L^2g_c'(1)}
\int^\infty_1 \frac{(y-1)y}{g_c}\left(\frac{kL^2}{r_{c}^2}+3+y(2+y)\right)
\frac{q^2\phi_c^2\psi_1^2}{g_c}dy. 
\end{align} 
Note that the first integral does not diverge according to the boundary condition 
at the horizon; $g_c'(1)\psi_1'(1)=m^2\psi_1(1)$. 

By demanding that the critical temperature in Eq.~(\ref{temperature}) is non-negative, 
we obtain an inequality $kL^2/r_{c}^2+6>0$, which guarantees the positivity of the 
second integral. So, by Eq.~(\ref{entropy}), we observe that the negative part of the effective 
mass~(\ref{effective-mass}), $q^2\phi_c^2/g_c$, increases the entropy of the hairly 
black hole at the second order $O(\epsilon^2)$. 

If the hairly black hole appears as the final fate of the instabilities found in Sec.~II, 
the entropy of the hairly black hole should be larger than the one of the RN-AdS$_4$ black 
hole according to the second law of the black hole. To check this, we numerically 
calculate the entropy difference between the hairly black hole and the RN-AdS$_4$ 
black hole: 
\begin{align}
\Delta S_2:=S_2|_{hairly BH}-{S_2}|_{RN-AdS_4 BH}
\end{align}
by integrating Eq.~(\ref{Delta-T}) for various values of $m^2$ and $k$.  
Since we are only interested in the sign of $\Delta S_2$, 
we numerically calculate $\Delta S_2$ for the solution of Eq.~(\ref{eq-psi1}) 
by normalizing as $\psi_1(1)=1$. 

\begin{figure}
\includegraphics[width=8.0truecm,clip]{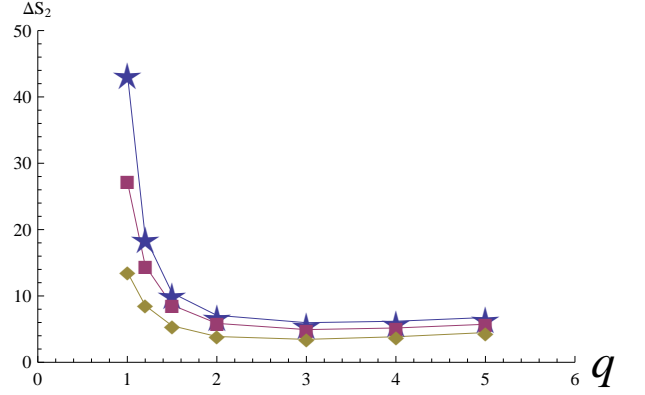}
\caption{ \label{fig:QNMs}
(color online). The $q$-dependence of $\Delta S_2$ for the solution of Eq.~(\ref{eq-psi1})
normalized as $\psi_1(1)=1$ in the case of $m^2=-2/L^2$, $\rho=2$. The star, box, and diamond 
represent $k=1$, $k=0$, and $k=-1$, respectively.}
\end{figure}
\begin{figure}
\includegraphics[width=8.0truecm,clip]{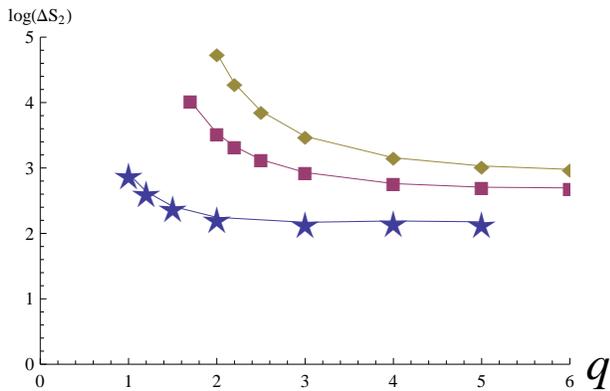}
\caption{ \label{fig:QNMs1}
(color online). The $q$-dependence of $\mbox{Log}_{10}(\Delta S_2)$ for the solution 
of Eq.~(\ref{eq-psi1}) normalized as $\psi_1(1)=1$ for various $m^2$ in the case of 
$k=0$, $\rho=60$. The star, box, and diamond represent $m^2=-2/L^2$, $m^2=0$, and 
$m^2=7/4L^2$, respectively. At the left edge of each plot, the RN-AdS$_4$ black holes 
become almost extremal.}
\end{figure}
In Fig.~3, we plot $\Delta S_2$ for various $q$ and $k$ in the case of 
$m^2=-2/L^2$ and $\rho=2$. Fig.~4 shows the $q$-dependence of 
$\mbox{Log}_{10}(\Delta S_2)$ for various $m^2$ in the case of $k=0$ and $\rho=60$. Both 
figures indicate that $\Delta S_2$ is positive, independent of the mass $m^2$, charge $q$, and 
curvature $k$. 

We also give a further evidence that the hairly black hole constructed above appears 
as the end state of the instabilities found in Sec.~II by calculating 
charge density $\tilde{\rho}$ outside the black hole horizon. 
$\tilde{\rho}$ on $T=const.$ hypersurface $\Sigma$ is calculated at $O(\epsilon)$ as
\begin{align}
\label{density-b}
\tilde{\rho}=j_an^a=2\epsilon\frac{q^2\phi_c(y)}{\sqrt{g_c(y)}}\psi_1^2=
\frac{2\epsilon q^2\rho\left(1-\frac{1}{y}\right)}{r_c^2\sqrt{g_c(y)}}\psi_1^2(y)>0,  
\end{align}
where $n^a$ is the past directed unit normal vector on $\Sigma$. So, positive 
charge is distributed outside the hairly black hole. This agrees with the 
result~(\ref{current}), where the positive charge is shown to be extracted from the horizon 
via the complex scalar field when RN-AdS$_4$ black hole becomes unstable.

\section{conclusion and discussions}
We have investigated instability of RN-AdS$_4$ black holes with various topologies against 
perturbation of charged scalar fields with various masses near the second order phase transition. 
Near the phase transition where the scalar field begins to condense, all fields are expected 
to evolve very slowly according to the critical slowing down~\cite{mno2008,mno2009}. 
So, we obtained a useful expression for the lowest QN frequencies of the RN-AdS$_4$ black 
holes by solving the scalar field equation~(\ref{eq-Psi}) perturbatively around the 
the phase transition. It is numerically shown that RN-AdS$_4$ black holes become unstable 
against the perturbations below the critical temperature $T_c$. 
By Eq.~(\ref{current}), we found that the charge extraction occurs from 
the RN-AdS$_4$ black holes during the unstable evolution.  

To see the final fate of the instabilities, we have also constructed the black hole solutions 
with the scalar hair by solving the field equations~(\ref{eq-motion}) perturbatively around the 
the phase transition. We found that the negative part of the effective mass ${\cal M}^2$ of the 
scalar field~(\ref{effective-mass}) increases the entropy.  
The numerical calculations indicate that the entropy of the hairly black holes is always 
larger than the one of RN-AdS$_4$ black holes in the microcanonical ensemble 
whenever RN-AdS$_4$ black holes are unstable. 
As indicated in Eqs.~(\ref{density-b}) and~(\ref{current}), the positive charge density 
outside the hairly black hole is extracted from the RN-AdS$_4$ black hole. These facts 
suggest that the hairly black holes are the final fate of the instabilities of RN-AdS$_4$ black 
holes, independent of the topology of the horizon and mass of the scalar field. So, 
this is a universal feature of the charged complex scalar system. 

Since the charge is extracted from the RN-AdS$_4$ black holes, the instability against 
the complex scalar field is very similar to the superradiance phenomenon where energy is  
extracted from a rotating black hole.   
The unstable modes inspired by superradiance were recently found in Kerr-AdS black 
hole~\cite{superradiance} only when the Einstein universe at the AdS boundary rotates faster 
than the speed of light. This is due to the fact that a global timelike killing vector field 
does not exist in the rapidly rotating case~\cite{hawking-real,kodama}. 
Refs.~\cite{hawking-real,kodama} also claim that there is no unstable mode for any 
linear perturbation equation when spacetime possesses a global timelike killing vector 
field $k^a=(\p_t)^a$ and the energy current $J^a:=-T_{ab}k^b$ satisfies the conservation 
law $\nabla_a J^a=0$ and the dominant energy condition.,~i.e.,~$J^a V_a\le 0$ for any 
future-directed timelike vector $V^a$. 

At first sight, this seems to contradict the unstable modes found in this paper,  
as they satisfy the dominant energy condition for $m^2\ge 0$ and 
RN-AdS$_4$ spacetime possesses the global timelike killing vector field $k^a$. 
However, the energy current $J^a$ of the scalar field is not conserved for the {\it fixed} gauge 
potential $A_a$ in Eq.~(\ref{eq:gauge-potential(0)}), as easily checked. 
This does not satisfy one of the conditions of the claim mentioned above.
Thus, energy can be transferred from the gauge field into the charged scalar field, 
like particle creation by an external field. 

In the asymptotically flat case, the superradiance scattering by a complex scalar 
field occurs for RN black holes and the reflected wave is amplified by the 
scattering potential~\cite{Gibbons75}. The condition for its occurrence depends on the sign 
of the gauge coupling $q$. In this sense, the instabilities found 
in Sec.~II is different from this usual picture of superradiance since  
$\omega_I$ in Eq.~(\ref{imaginary-omega}) does not depend on the sign 
of $q$. A detailed analysis would be necessary in this direction. 

In this article, we restrict our interest into the pure-electric case. 
As shown in Ref.~\cite{mno2010}, the plane-symmetric RN-AdS$_4$ possesses an inhomogeneous 
scalar hair with a vortex lattice structure under the probe approximation when magnetic field is 
included. It is one of interesting directions of our research to investigate what 
happens when the back-reaction of the inhomogeneous scalar field onto the geometry.

\begin{acknowledgments} 
We would like to thank Makoto Natsuume, Takashi Okamura, Hideo Kodama for useful discussions.
This research was supported in part by the Grant-in-Aid for Scientific
Research~(20540285) from the Ministry of Education, Culture,
Sports, Science and Technology, Japan.
\end{acknowledgments}

\appendix 
\section{The first law of black hole thermodynamics} 

In this appendix, we verify that the first law of thermodynamics 
holds for the asymptotically anti-de Sitter black hole with the complex 
scalar hair $\psi$, for any value of $k$ of the 2-dimensional Einstein space.  
To do so, we follow the method developed by 
Wald and collaborators \cite{SympFirstLaw}. Although it can be 
applicable to rotating black holes, for simplicity, we focus on the static black 
holes with the 2-dimensional Einstein space of constant curvature.    

We start with the variation of 
$\varepsilon_{c_1 c_2 c_3 c_4} \mathcal{L}(g_{a b}, A_a , \psi)$, 
which is written as  
\begin{equation}
\delta \left[ \varepsilon_{c_1 c_2 c_3 c_4} \mathcal{L}(\mathcal{F}) \right] = 
\varepsilon_{c_1 c_2 c_3 c_4} \: 
\nabla_b \: \Theta^b(\mathcal{F}, \delta \mathcal{F}) , 
\label{eqn:ActionVariation}
\end{equation} 
where $\varepsilon_{c_1 c_2 c_3 c_4}$ is the volume element of the spacetime, 
$\mathcal{F}$ denotes the field variables $g_{a b}$, $A_a$, and $\psi$, collectively.  
$\Theta^b(\mathcal{F}, \delta \mathcal{F})$ in Eq. (\ref{eqn:ActionVariation}) is 
derived for the Lagrangian density (\ref{lagrangian-density}) as 
\begin{equation}
\Theta^b(\mathcal{F}, \delta \mathcal{F}) 
= \Theta_{(g)}^b(\mathcal{F}, \delta \mathcal{F}) 
+ \Theta_{(A)}^b(\mathcal{F}, \delta \mathcal{F}) 
+ \Theta_{(\psi)}^b(\mathcal{F}, \delta \mathcal{F}) , 
\end{equation} 
where 
\begin{align} 
\Theta_{(g)}^b(\mathcal{F}, \delta \mathcal{F})  & 
= g_{c d} \nabla^b \delta g^{c d} - \nabla_c \delta g^{b c} , 
\\ 
\Theta_{(A)}^b(\mathcal{F}, \delta \mathcal{F}) & 
= - F^{b c} \delta A_c , 
\\
\Theta_{(\psi)}^b(\mathcal{F}, \delta \mathcal{F}) & 
= - ( D^b \psi ) \, \delta \psi^{\dagger} 
- ( D^b \psi)^{\dagger} \delta \psi . 
\end{align} 

When the variation $\delta$ is given by infinitesimal diffeomorphism, i.e., 
the Lie derivative $L_{\eta}$ along an arbitrary vector $\eta^a$, 
the current $J^b(\mathcal{F}, \eta)$ defined by 
\begin{equation}
J^b(\mathcal{F}, \eta) \equiv 
\Theta^b(\mathcal{F}, L_{\eta} \mathcal{F}) - \eta^b \: \mathcal{L}(\mathcal{F})  
\label{eqn:JDef}
\end{equation}
is conserved on shell, and we can find the Noether potential $Q^{b a}(\mathcal{F}, \eta)$ 
of $J^b(\mathcal{F}, \eta)$, such that 
\begin{equation}
J^b(\mathcal{F}, \eta) \equiv \nabla_a Q^{b a}(\mathcal{F}, \eta) . 
\label{eqn:QDef}
\end{equation} 
For the Lagrangian density (\ref{lagrangian-density}), 
$Q^{b a}(\mathcal{F}, \eta)$ is derived as 
\begin{equation}
Q^{b a}(\mathcal{F}, \eta) = Q_{(g)}^{b a}(\mathcal{F}, \eta) 
+ Q_{(A)}^{b a}(\mathcal{F}, \eta) , 
\end{equation} 
where 
\begin{align} 
Q_{(g)}^{b a}(\mathcal{F}, \eta) & 
= \nabla^a \eta^b - \nabla^b \eta^a , 
\\ 
Q_{(A)}^{b a}(\mathcal{F}, \eta) & 
= F^{a b} A_c \eta^c ,  
\end{align} 
and the contribution to $Q^{b a}(\mathcal{F}, \eta)$ from the scalar field 
is absent \cite{JacobsonKM}. 

In particular, when $\eta^a$ is a Killing vector $\xi^a$, 
one can show 
\begin{equation} 
0 = \delta \int_{\partial \Sigma} \frac{1}{2} \; 
\varepsilon_{b a c d} \; Q^{b a}(\mathcal{F}, \xi)  
+ \int_{\partial \Sigma} \varepsilon_{b a c d} \; 
\xi^b \Theta^a(\mathcal{F}, \delta \mathcal{F}) ,  
\label{eqn:HamiltonianSurface} 
\end{equation} 
for an arbitrary hypersurface $\Sigma$. 
In order to derive the first law, we choose as $\xi^a$ the timelike Killing vector 
$k^a$ that becomes null at the horizon, and 
$\Sigma$ to be a spacelike hypersurface connecting the spacial infinity and 
the bifurcation surface $\mathcal{B}$ of the black hole horizon, which we 
assume to exist. 

Thus, the boundary $\partial \Sigma$ includes the whole (in the case of $k = 1$) or 
a portion (for $k = 0$ and $k = - 1$) of $\mathcal{B}$ and 
the 2-dimensional Einstein space at infinity. In addition, in the case of 
$k = 0$ or $k = - 1$, there are other parts of the boundary, because 
the 2-dimensional Einstein space is not closed. 
However, in this case, we respect the isometries of the 2-dimensional Einstein space 
in choosing $\Sigma$, and impose a periodic boundary condition at these parts of 
the boundary. Then, the contribution from these parts of the boundary 
vanishes in Eq. (\ref{eqn:HamiltonianSurface}).

We define 
the total mass (for $k = 1$) or the mass density 
(for $k = 0$ and $k = -1$) of the black hole, which we denote as $M$, by  
\begin{equation}
M \equiv \frac{1}{4 w_2} \int_{\infty} \frac{1}{2} \; 
\varepsilon_{b a c d} \left[ Q_{(g)}^{b a}(\mathcal{F}, k)  
+ 2 \: k^{[ b} B^{a ]}(\mathcal{F}) \right] ,  
\label{eqn:MassDef} 
\end{equation} 
where $B^a(\mathcal{F})$ is defined by 
\begin{equation}
\delta \int_{\infty} \varepsilon_{b a c d} \; 
k^b B^a(\mathcal{F}) 
= \int_{\infty} \varepsilon_{b a c d} \; 
k^b \Theta_{(g)}^a(\mathcal{F}, \delta \mathcal{F}) , 
\end{equation} 
and $w_2$ is the area of the whole (for $k = 1$) or the portion (for $k = 0$ and $k = -1$) 
of the 2-dimensional Einstein space with the metric $d\Omega_2^2$. 
It has been shown \cite{HollandsIM05} 
that the mass defined by Eq. (\ref{eqn:MassDef}) coincides with 
other various definitions of mass.
The electric charge~(or charge density) $\rho$ of the black hole is defined as 
\begin{equation}
\rho \equiv \frac{1}{w_2} \int_{\infty} \frac{1}{2} \; 
\varepsilon_{b a c d} F^{b a} , 
\end{equation} 
and the static potential is given by $- k^a A_a$, as usual. 
The gauge invariant quantity associated with the static potential, 
in the first law of black hole thermodynamics, 
is the difference of its value between the horizon and infinity.  
Since $k^a = 0$ on $\mathcal{B}$, this quantity is shown, in a regular gauge, to coincide 
with the chemical potential $\mu$ of the field theory, due to AdS/CFT, as 
\begin{equation}
\left( - \left. k^a A_a \right|_{\mathcal{B}} \right) 
- \left( - \left. k^a A_a \right|_{\infty} \right) 
= \left. k^a A_a \right|_{\infty} = \mu  . 
\end{equation} 
Furthermore, for the boundary condition Eq. (\ref{bn-Dirichlet}) of the scalar field, 
$\psi \sim r^{- \Delta_+}$, and we have 
\begin{equation}
\int_{\infty} \varepsilon_{b a c d} \; 
k^b \Theta_{(\psi)}^a(\mathcal{F}, \delta \mathcal{F}) = O(r^{- 2 \Delta_+ + 3}) , 
\end{equation}
which shows that the contribution of the scalar field at infinity vanishes 
since $\Delta_+>3/2$. 

Therefore, by defining the entropy~(or entropy density) $S$ of the black hole as 
\begin{equation}
T S \equiv \frac{1}{4 w_2} \int_{\mathcal{B}} \frac{1}{2} \; 
\varepsilon_{b a c d} \; Q^{b a}(\mathcal{F}, k) ,  
\end{equation}
and noting $k^{[ a} F^{b c ]} = 0$ and the fact 
that $k^a$ vanishes on $\mathcal{B}$, we obtain the first law as 
\begin{equation}
T \delta S = \delta M - \frac{\mu}{4} \, \delta \rho . 
\label{eqn:FirstLaw} 
\end{equation}
While one can explicitly derive Eq. (\ref{eqn:FirstLaw}) for the RN-AdS$_{4}$ black hole, 
from Eqs. (\ref{temperature}), (\ref{def-rho-mass}), and (\ref{eq:gauge-potential(0)}), 
it is satisfied also for hairy black holes.

\end{document}